\documentclass[
    ,final            
  ]
  {aipproc}

\layoutstyle{6x9}

\usepackage{color}
\begin{document}

\title{Mixing by Non-linear Gravity Wave Breaking on a White Dwarf Surface}

\author{A.~C.~Calder}{
  address={Center for Astrophysical Thermonuclear Flashes,
                 The University of Chicago,
                 Chicago, IL  60637},
  altaddress={Department of Astronomy and Astrophysics,
                 The University of Chicago,
                 Chicago, IL  60637},
  email={calder@flash.uchicago.edu}
}

\author{A.~Alexakis}{
  address={Department of Physics,
                The University of Chicago,
                 Chicago, IL  60637}
}

\author{L.~J.~Dursi}{
  address={Center for Astrophysical Thermonuclear Flashes,
                 The University of Chicago,
                 Chicago, IL  60637},
  altaddress={Department of Astronomy and Astrophysics,
                 The University of Chicago,
                 Chicago, IL  60637}
}

\author{R.~Rosner}{
  address={Center for Astrophysical Thermonuclear Flashes,
                 The University of Chicago,
                 Chicago, IL  60637},
  altaddress={Department of Astronomy and Astrophysics,
                 The University of Chicago,
                 Chicago, IL  60637},
}

\author{J.~W.~Truran}{
  address={Center for Astrophysical Thermonuclear Flashes,
                 The University of Chicago,
                 Chicago, IL  60637},
  altaddress={Department of Astronomy and Astrophysics,
                 The University of Chicago,
                 Chicago, IL  60637}
}

\author{B.~Fryxell}{
  address={Center for Astrophysical Thermonuclear Flashes,
                 The University of Chicago,
                 Chicago, IL  60637},
  altaddress={Enrico Fermi Institute,
                 The University of Chicago,
                 Chicago, IL  60637}
}

\author{P.~Ricker}{
  address={Center for Astrophysical Thermonuclear Flashes,
                 The University of Chicago,
                 Chicago, IL  60637},
  altaddress={Department of Astronomy and Astrophysics,
                 The University of Chicago,
                 Chicago, IL  60637}
}

\author{M.~Zingale}{
  address={Center for Astrophysical Thermonuclear Flashes,
                 The University of Chicago,
                 Chicago, IL  60637},
  altaddress={Department of Astronomy and Astrophysics,
                 The University of California,
                 Santa Cruz, CA  95064}
}

\author{K.~Olson}{
  address={UMBC/GEST Center, 
                NASA/GSFC, Greenbelt, MD 20771}
}

\author{F.~X.~Timmes}{
  address={Center for Astrophysical Thermonuclear Flashes,
                 The University of Chicago,
                 Chicago, IL  60637},
  altaddress={Department of Astronomy and Astrophysics,
                 The University of Chicago,
                 Chicago, IL  60637}
}

\author{P.~MacNeice}{
  address={UMBC/GEST Center, NASA/GSFC, Greenbelt, MD 20771}
}

\begin{abstract}
We present the results of a simulation of a wind-driven non-linear gravity wave 
breaking on the surface of a white dwarf. The ``wind'' consists of H/He from an
accreted envelope, and the simulation demonstrates that
this breaking wave mechanism can produce a well-mixed layer of H/He with
C/O from the white dwarf above the surface. Material from this
mixed layer may then be transported throughout the accreted envelope by
convection, which would enrich the C/O abundance of the envelope as is
expected from observations of novae.
\end{abstract}

\maketitle


\section{Introduction}
Classical novae result from the ignition (and subsequent explosive
thermonuclear burning) of a ($\sim 10^4$ m) layer of hydrogen-rich material that 
has accreted from a main sequence companion onto the surface of a white
dwarf \cite{truran82,shara89,starrfield89,livio94}.
Observed abundances and explosion energies estimated from observations
indicate that there must be significant mixing of the heavier material
of the C/O or O/Ne white dwarf into the lighter accreted material
(H/He). This mixing is critical because otherwise hydrogen burning
would be too slow to reproduce observed nova characteristics in
outburst. Further, without this mixing it is difficult to understand
the observed abundances of intermediate-mass nuclei in the ejecta.
Accordingly, nova models must incorporate a mechanism that will dredge up
the heavier white dwarf material~\cite[and references therein]{rosner01}.

A recently proposed mixing mechanism is the breaking of non-linear resonant
gravity waves at the C/O surface \cite{rosner00,alexakis01,rosner01}.
The gravity waves, driven by the ``wind'' of accreted material, can break,
forming a layer of  well-mixed material. This mixed layer may then
be transported upward by convection, thereby enriching the accreted material.
Because the length scale of this mixed layer may be very small (much
smaller than the length scale of convection), previous precursor
simulations have not captured this effect.

In this manuscript, we present a simulation of a wind-driven non-linear 
gravity wave breaking on the surface of a white dwarf. The simulation was 
performed with FLASH, a parallel, adaptive-mesh simulation code for the 
compressible, reactive flows found in many astrophysical 
environments~\cite{fryxell00,calder00}. This simulation is part of an
ongoing study of this mechanism to assess its efficacy for mixing white
dwarf material with envelope material.
\begin{figure}
  \includegraphics[height=.3\textheight]{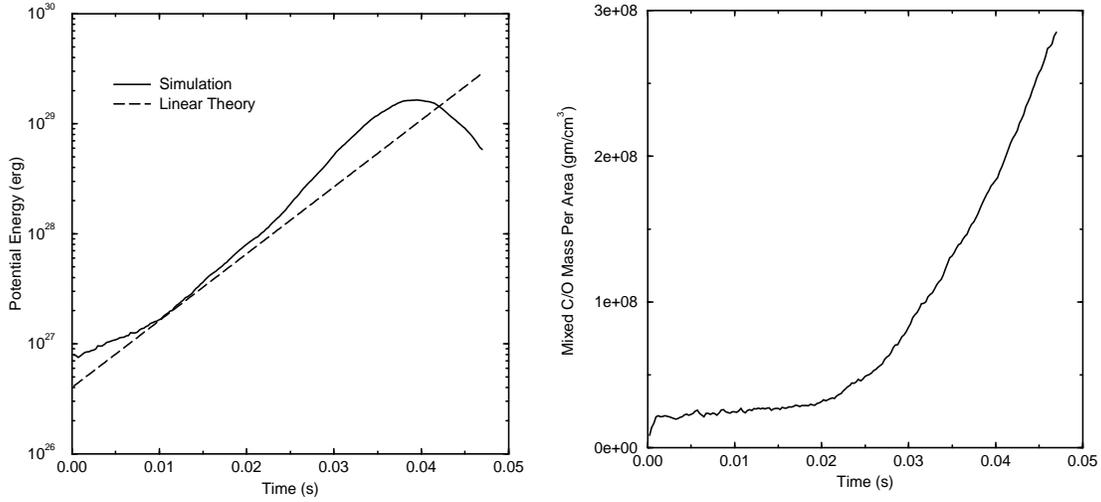}
  \caption{Results from the simulation of a wind-driven gravity wave. The left
panel shows the potential energy of the wave vs.\ time. Also
shown is the potential energy predicted by the linear theory.
The right panel shows the mixed C/O mass per unit area vs.\ time.
\label{fig:mixpe}}
\end{figure}

\section{Theory and Simulation Details}

This work applies the theory of gravity wave generation (originally 
developed for the air over water interface)~\cite[and references 
therein]{miles57} to a white dwarf with an 
accreted envelope. The control parameters for this model
are $G = g\delta/U_{\rm max}^2$, which is the ratio of potential to
kinetic energy 
of the wind ($g$ is the constant gravitational acceleration,
$U_{\rm max}$ is the maximum wind speed, and $\delta$ is the characteristic 
length scale of the wind profile) and the ratio of densities
at the interface, $r = \rho_1/\rho_2$.
For this simulation, the domain was $1.0 \times 10^6 $ cm by $1.0 \times 10^6 $ cm,
$g = 4.5 \times 10^9$ cm/s, and $r = 10$, with a density of the
white dwarf material at the interface of $10^4$ gm/cm$^3$.
The white dwarf and accreted envelope materials are modeled as 
simple $\gamma = 5/3$ gases
with the white dwarf material composed (by mass) of a 50/50 C/O mix and
the accreted material composed of a 75/25 H/He mix.

The density and pressure profiles were obtained by integrating the 
equation of hydrostatic equilibrium
\begin{equation} \label{eq:hydrostat} 
\frac{dp}{dy} = -\rho g {\bf \hat{k}} \;,
\end{equation} 
which for the case of a compressible, gamma-law gas gives
\begin{equation} \label{eq:compr}
\rho = \rho_i\left[1 -\left(\gamma
- 1\right)\frac{g \rho_iy}{P_0 \gamma} \right]^{\frac{1}{\gamma - 1}}
\end{equation}
and
\begin{equation} \label{eq:compp}
P = P_0 \left[1 -\left(\gamma - 1\right)\frac{g \rho_iy}{P_0 \gamma}
\right]^{\frac{\gamma}{\gamma - 1}} \; .
\end{equation}
Here $P_0$ is the pressure at the interface and $\rho_i$ 
is the density immediately above or below the interface.
The wave was created by forcing the interface
to be sinusoidal and perturbing the pressure and adding a velocity via
a prescription similar to linear theory.
\begin{figure}
  \includegraphics[height=.27\textheight]{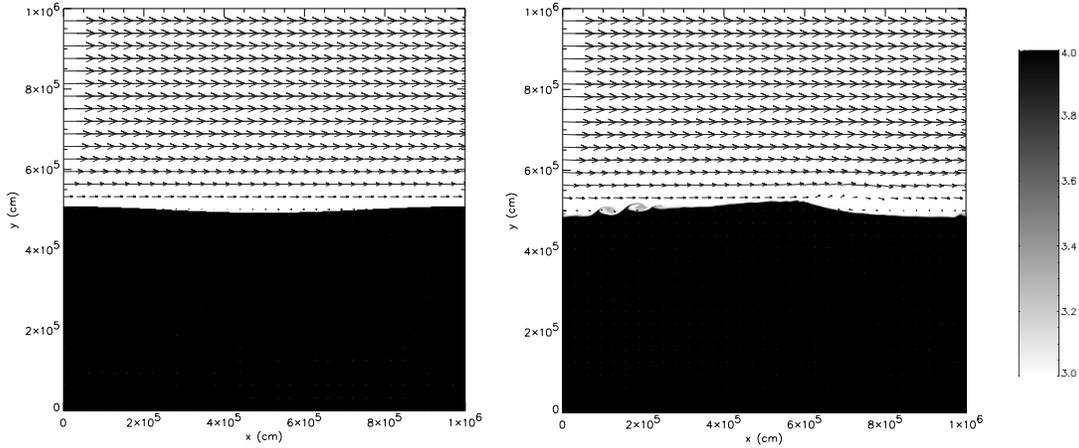}
  \caption{Images of the log of density with velocity vectors at earlier 
times during the simulation. The units of density are g/cm$^3$ and 
the length of the velocity arrows is proportional to the magnitude
of the velocity, with a maximum of $2 \times 10^8$ cm/s.
The left panel shows the initial 
conditions, and the right panel shows the simulation at t = 0.015 s.
\label {fig:early}}
\end{figure}
The wind profile is given by
\begin{equation} \label{eq:windp} 
U(y) = U_{\rm max} \left(1 - e^{-y/ \delta}\right)
\end{equation} 
with $\delta = 1.0 \times 10^5$ cm 
and 
$U_{\rm max} = 2 \times 10^8~{\mathrm{cm/s}}$.  The boundary conditions were
isothermal hydrostatic on the upper and lower boundaries and periodic on 
the sides.
The simulation was performed with seven levels of adaptive mesh refinement,
for an effective resolution of $512 \times 512$ zones.

\section{Results}

Figure~\ref{fig:mixpe} shows the potential energy of the wave 
and the amount of mixed material 
during the course of the simulation. The left panel is a plot
of potential energy vs.\ time showing the result from the simulation 
and the potential energy given by the
linear theory as $4.0 \times 10^{26} {\rm exp}(140t)$. The growth rate
in this expression (140) comes from the control parameter $G$, 
and $4.0 \times 10^{26}$ (erg) is the initial potential energy of the
wave~\cite{alexakis01}.
The two potential energies agree reasonably well until the wave breaks
at about 0.04 s.
The right panel is a plot of mixed C/O mass per unit area vs.\ time.
The figure demonstrates the dramatic increase in the amount of
mixed material that occurs above $t = 0.25$, the point during the course
of the simulation at which the wave begins to break.

\begin{figure}
  \includegraphics[height=.27\textheight]{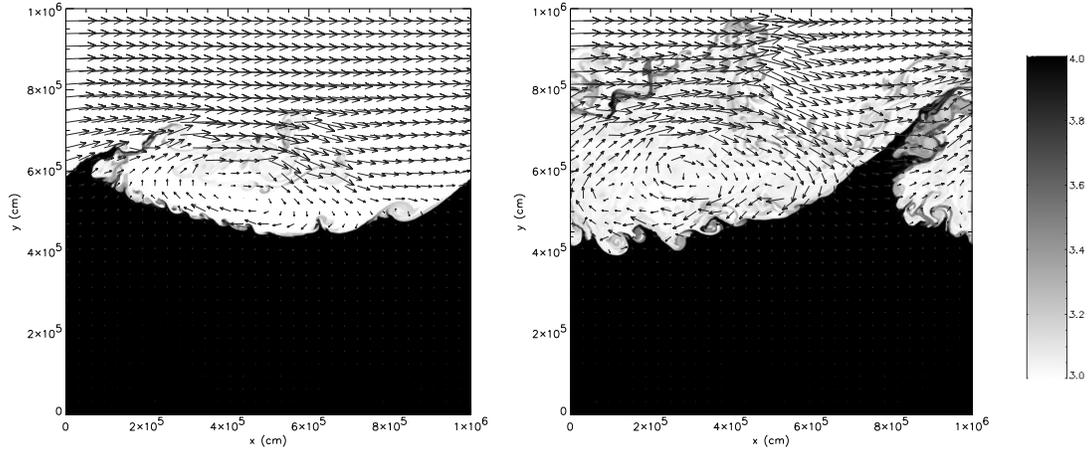}
  \caption{Images of the log of density with velocity vectors at later 
times during the simulation. The units of density are g/cm$^3$ and 
the length of the velocity arrows is proportional to the magnitude
of the velocity, with a maximum of $2 \times 10^8$ cm/s.
The left panel shows the simulation at
t = 0.030 s, and the right panel shows the simulation at t = 0.045 s.
\label{fig:late}}
\end{figure}

Figure~\ref{fig:early} shows gray scale images of the log of density with 
velocity vectors at early times in the simulation. The longest 
velocity arrows, at the top of the images,
correspond to $2.0 \times 10^8$ cm/s, the maximum wind speed. 
The left panel shows the initial 
conditions, and the right panel shows the configuration at $t$ = 0.015 s.
Visible in the right panel is the development of Kelvin-Helmholtz instability
on the upwind (left) side of the wave.
Figure~\ref{fig:late} shows similar gray scale images of the 
log of density with velocity vectors at later times in the simulation. 
By $t = 0.030$ s (left panel), the wave
has begun to break, and the images show vorticity downwind
from the crest of the wave. At $t = 0.045$ s (right panel), there is 
substantial mixing and obvious vorticity.

\section{Conclusions}

The simulation presented in this manuscript demonstrates the
proposed breaking wind-driven non-linear gravity wave mixing mechanism.
The results show the development of well-mixed zone just above the
surface of the white dwarf. This simulation will be one part of
a study of this mixing mechanism investigating effects of wind
profiles and speeds. Complete details of the study 
will appear in \citet{alexakis02}. 

The expectation is that the breaking of non-linear gravity waves
on the surface of the white dwarf will lead to a thin well-mixed
layer of material that may then be transported throughout the
the envelope by convection. This study investigating the mixing
mechanism should provide quantitative
information about the mixing rate that will allow for the
development of subgrid models that may be applied to multidimensional
convection simulations to study the enrichment of the envelope.
A preliminary simulation of this kind is also presented in this 
volume ~\cite{dursi02}.

\begin{theacknowledgments}
This work is supported in part by the U.S. Department of Energy (DOE) under
Grant No. B341495 to the Center for Astrophysical Thermonuclear Flashes
at the University of Chicago. J.~W.~Truran acknowledges partial support
from DOE grant DE-FG02-91ER40606.  L.~J.~Dursi is 
supported by the Krell Institute CSGF. K.~Olson acknowledges partial 
support from NASA grant NAS5-28524. M. Zingale acknowledges support 
from the Scientific Discovery through Advanced Computing
(SciDAC) program of the DOE, grant number DE-FC02-01ER41176.
Additional details about the project and information about
requesting a copy of FLASH may be found at \url{http://flash.uchicago.edu}.

\end{theacknowledgments}

\end{document}